\documentclass{article}
\usepackage{amsmath,amssymb,amsfonts}
\usepackage{algorithmic}
\usepackage{graphicx}

\begin{document}

\title{On Aadhaar Identity Management System}

\author{Yash Mehta, Dev Patel, and Manik Lal Das\\
DA-IICT\\
Gandhinagar, India.}

\date{}

\maketitle

\begin{abstract}
A unique identification for citizens can lead to effective
governance to manage and provide citizen-centric services. While
ensuring this service, privacy of the citizens needs to be
preserved. Aadhaar, the identification system by UIDAI has faced
some critics regarding its privacy preserving feature. This paper
discusses those concerns in Aadhaar system and proposed a new
model for the Aadhaar system. The proposed solution is aimed to
address the issue of collusion of third party service providers
and profiling of Aadhaar users. The proposed solution uses a
distributed model capturing the Aadhaar system, in which data of
users is decentralized and stored in zonal office's databases as
well as the CIDR. The proposed solution provides the functioning
of the authentication process of the Aadhaar system more
effective, as it reduces the number of requests being handled
directly by the CIDR and also tackles the concern of correlation
of data.\vspace{1 mm}\\
\textbf{Keywords.} Aadhaar system; Authentication; Privacy;
Distributed System.
\end{abstract}

\section{Introduction}
A unique identification for all citizens has become a necessity
for any Government for availing various services offered to
citizens of the country. In a country like India, where a large
population comes under common subsidies offered in different
sectors, the unique identification can be used for direct transfer
of subsidies to the citizens, which brings transparency,
accountability and efficacy in the system. When the verification
of the citizens is required, a centralized single identification
number can help speed up the process of verification and the need
of multiple documents for verification is eliminated. This greatly
reduces time and effort required to avail services. Therefore, the
benefits of such an identification number are enormous. Many
countries have been maintaining such a system since quite long.
The Aadhaar, a unique identification number \cite{aadhaar act} by
Government of India enables all
citizens of the country under one identification system.\\
Aadhaar is a 12-digit unique number assigned by the Unique
Identification Authority of India (UIDAI) to the citizens of the
country after satisfying the verification process laid down by the
Authority. To enrol for the Aadhaar, the person needs to visit an
enrollment centre, fill the enrolment form, get biometric data,
including ten fingerprints, two iris scans and a photograph,
captured in the devices provided, and submit proof of identity and
address documents. All these data are stored in the Central
Information Data Repository (CIDR). Therefore, a huge database of
individuals sensitive information is to be maintained. The While
Aadhaar has its numerous benefits, the Aadhaar system has faced
some criticisms for its privacy preserving features. Many
questions have been raised concerning the security of the private
information of the users. The problems with the initial
architecture of the Aadhaar came to light stating that the private
data of the users got leaked \cite{toi}. Details including name,
address, photo, phone number and email address were available.
Data leaks are always a threat when dealing with a large and
sensitive database. Furthermore, as Aadhaar is used for
authentication by various third party service providers, these
service providers could collude amongst themselves and profile a
user, which is an invasion of the privacy of people. While some
research works have tried to address the problems faced in
Aadhaar, we present a different approach to address the same
issues. A distributed approach is considered by making
modifications to the Aadhaar system. The proposed architecture is
using cluster-based data decentralization notion to make the
Aadhaar system a secure system by addressing the concern of
correlation of data and user profiling. The remaining of the paper
is structured as follows. In section 2, we discuss some related
works. In section 3, the proposed solution is presented. In
section 4, we provide the analysis of the proposed solution. We
conclude the paper in section 5.

\section{Related Work}

\subsection{Existing Operation Model of Aadhaar Authentication}
The Figure 1 describes how the different entities involved in
Aadhaar system interact with each other.

\begin{figure}[h!]
  \centering
  \includegraphics[height=5 cm, width=\linewidth]{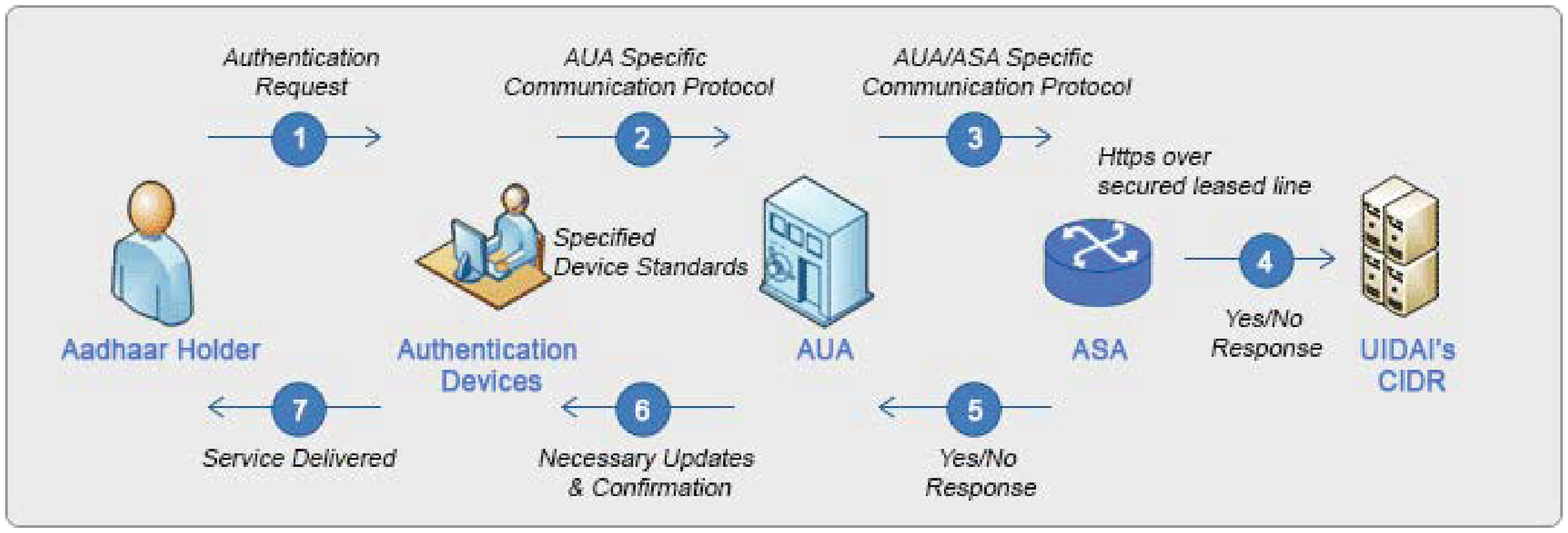}
  \caption{System Architecture of Aadhaar system \cite{aadhaar dashboard}}
\end{figure}
\noindent The authentication process works as follows.

\begin{itemize}
\item The Aadhaar user gives the Aadhaar number and necessary
biometric, demographic data and/or OTP as input to the client
application. The client application packages and encrypts these
input parameters into a PID block before any transmission, as per
authority standards and sends it to the server of the requesting
entity using secure protocols laid down by the authority.

\item After validation, the server forwards the authentication
request to the CIDR through the server of ASA as per authority
specifications. This request is digitally signed by the requesting
entity and/or ASA.

\item The CIDR validates the input parameters against the data
stored in the CIDR and returns a digitally signed Yes/No response
or a digitally signed e-KYC authentication response with encrypted
e-KYC data, along with other technical details related to the
authentication transaction.
\end{itemize}
It is noted that the privacy of the data of the users, misuse of
authentication without consent of the user are important concerns.
Research and observations \cite{rajput} \cite{anusha}
\cite{shweta} highlighted security and privacy concerns in Aadhaar
concerns. For better readability, we elaborate these concerns in
detail and also the solutions suggested by research community.

\subsubsection{Data Leaks}
The CIDR stores private data of users including their biometrics
and demographics. While this data may seem harmless, it can be
misused in many ways like stealing the identity of the actual
user. Incidences of data leaks of Aadhaar have come to light with
the sensitive data of users. The UIDAI collects data using
end-to-end encryption, thus preventing data leaks in transit
whenever a new user enrols for the Aadhaar. However, potential
threat that an employee or an intermediary with access to the CIDR
data may still leaks data of Aadhaar users.

\subsubsection{Authentication without Consent}
Aadhaar stores biometric details of the users such as
fingerprints, iris and a facial photograph. Fingerprints can
easily be obtained from any household object touched. With the
advent of technology, iris scans can also be obtained from a
photograph of the person. As a consequence, an unknown person can
identify and act as proxy authenticate pretending as some other
person whose biometrics he/she has obtained illegally, which could
result into serious crimes such as identity thefts. It is
therefore necessary to ensure that the authentication process is
started by the actual user in Aadhaar system.

\subsubsection{Correlation of Data}
A user may avail different services offered by service providers
by using the Aadhaar number for verification purposes. These
service providers could collude amongst themselves and gather
different data about the user. This way, they can profile a user
by correlating the data gathered. Once the profiles are generated,
this information can be used in malicious ways. Facebook-Cambridge
Analytica data leakage and data correlation is a good example
\cite{harry}, \cite{nieva}to see how correlation of data can
become a potential threat in Aadhaar system.\\
UIDAI has addressed this concern by enforcing service providers to
use local ids and maintain a mapping from Aadhaar number to the
local id. This, in our views, works to certain degree of trust
assumption; however, the service providers still can have access
to Aadhaar numbers and can track and profile users. A possible way
out is to maintain a unidirectional reverse linking from the local
ids to the Aadhaar numbers. Nevertheless, this solution also is
not enough because if the linking is stored with service
providers, they have indirect access to the Aadhaar numbers.
Rajput and Gopinath \cite{rajput} proposed another solution to the
problem of correlation. Their solution requires the user to
initiate the authentication process, thus preventing
identification without consent. A third party provides a temporary
id to the user requesting authentication and further
authentication process is done using the temporary id. As a
result, no service providers have access to the Aadhaar number of
the user and hence can not collude amongst themselves to profile
any user. However, it is to be noted that to generate the
temporary id, the Aadhaar user does need to give his/her Aadhaar
number to the third party generating the temporary id using which
the third party maps the id to the Aadhaar number. Again, the
question pops up. What if the third party colludes with some
service provider? The assumption made in \cite{rajput} is the
third party is trustworthy, but a big question is can we have this
hold true across the board in practice?

\section{The Proposed Solution}
We present a modified architecture for Aadhaar system with the
basic aim to prevent correlation of data. We primarily restrict
third party involvement in user's Aadhaar number verification.
Furthermore, the service provider only receives a Yes/No response
from the UIDAI as far the Aadhaar number verification is
concerned.

\subsection{Proposed System Architecture}
A decentralized approach is required, where UIDAI can divide the
data processing task into different zones in the country and
develops zonal offices in each zone. The zonal offices store
Aadhaar card details of the users who live in their zones. A minor
extension is required in the enrollment process due to this
change. After enrollment of each new user and the storage of data
in the CIDR, the same data needs to be stored in the concerned
zonal office as well. The Figure 2 represents the proposed system
architecture.

\begin{figure}[h!]
  \centering
  \includegraphics[height=5 cm, width=\linewidth]{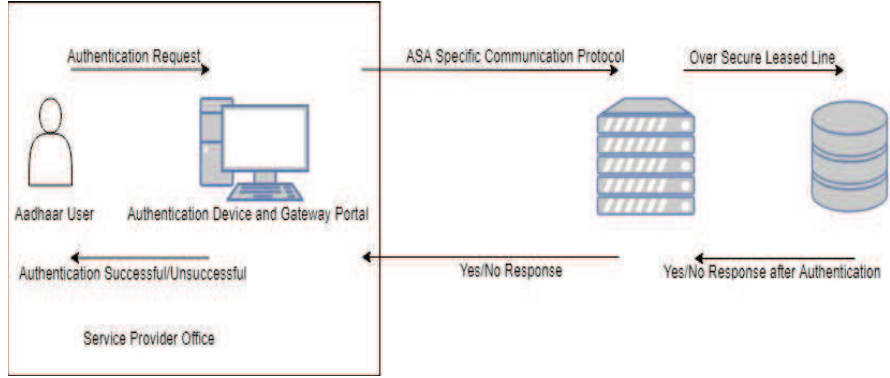}
  \caption{The Proposed System Architecture}
\end{figure}
\noindent In precise terms, we make the following modifications to
the existing operational model:

\begin{itemize}
\item The role of the service provider gets reduced in the
verification process.

\item The CIDR is replaced by the UIDAI zonal office. A database
containing the data of users living in that particular zone is
stored in the zonal office.
\end{itemize}

\subsection{Authentication Process}
The authentication process works in the following way:

\newcounter{1}
\begin{list}{\arabic{1}}
{\usecounter{1}} \item The service provider gives access to a
secure gateway portal provided by the zonal office of the
respective zone.

\item The user provides all the information required for
verification on this portal. This information is first packaged
into a block and encrypted.

\item This package is forwarded to the zonal office through the
server of ASA as per the specifications.

\item The zonal office then performs the authentication process by
comparing the data provided by the user with the data stored
beforehand.

\item If the information provided by the user is correct, then the
portal transmits a digitally signed Yes/No response to the service
provider over the secure line through ASA. This way the service
provider can verify the user without getting any information about
the user.

\item If the information of the user does not exist in the zonal
database, the zonal office sends a request to the CIDR to forward
the information of the user by providing the Aadhaar number. The
information received from the CIDR is to be stored only in cache
memory and hence, is discarded later on. Thus, only the data of
users in the pertaining zone remain in persistent storage.
\end{list}
\noindent The following algorithm explains the authentication
process.

\begin{algorithmic}
    \STATE Receive Package from ASA Server
    \STATE Validate Signature and extract Aadhaar Number
    \IF{PID data for Aadhaar number found}
        \STATE Extract User Submitted Data
        \IF{Stored PID Data matches User Submitted Data}
            \STATE ``Authentication Successful''
        \ELSE
            \STATE ``Authentication Unsuccessful''
        \ENDIF
    \ELSE
    \STATE Send Request to CIDR for PID
        \IF{PID data received}
            \STATE Extract User Submitted Data
             \IF{Stored PID Data matches User Submitted Data}
                \STATE ``Authentication Successful''
            \ELSE
                \STATE ``Authentication Unsuccessful''
            \ENDIF
        \ELSE
            \STATE ``Invalid Aadhaar Number. Please try Again...''
        \ENDIF
    \ENDIF
    \STATE Authentication Process Complete\\
\end{algorithmic}
\vspace{2 mm}\noindent It is noted that during the authentication
process, no private information of the users is accessible to any
third party or service providers. The proposed architecture
reduces the traffic at the CIDR by distributing the data among
different zonal offices.

\section{Analysis of the Proposed Solution}

\subsection{Prevention of unauthorized consent}
One of the major concerns with the Aadhaar system is that
biometric data like fingerprints of a person can be easily
obtained from the objects touched by that person. Iris scans can
also be obtained from high resolution photos. Hardware and
software that supports liveness detection could avoid this
problem. Using this approach the hardware issued by the zonal
office or legitimate authority should check whether a living
person is providing the biometric data and iris scans running some
software. Therefore, the user will be able to provide the
biometric and demographic data and that too on designated devices
so that no other person will be able to retrieve any data
pertaining to user authentication and verification.

\subsection{Prevention of correlation and profiling}
We have discussed how third party service providers can collude
together to profile a person and can use the data with malicious
intention. In the proposed solution, when the third party service
provider asks the Aadhaar number and data for the verification the
user is directed to a portal provided by the authorized zonal
office. The Aadhaar number and other data pertaining to the user
is then uploaded to that portal and then the verification is done
by the zonal office/servicer. As the verification is avoiding the
involvement of third part service providers, data correlation and
user profiling can be prevented in the proposed solution.

\subsection{Security and Efficiency}
In the proposed solution, the data and tasks on data, both are
segregated and clustered in different zones based on the users'
demography and preference. The user data is stored at zonal
offices without involving any third party, so the data is stored
and managed with better control and security. In case of a new
entry, minimal operation as well as update is required in managing
the data. Importantly, the search operation is performed locally,
which is a frequent operation and the proposed solution
facilitates better efficiency for Aadhaar system.

\section{Conclusion}
We have discussed the Aadhaar system and observed some concerns in
the system. The system should ensure that authentication with the
live involvement of the user in the system. Profiling of users by
correlation of data due to collusion by third party service
providers needs to be prevented in the system. The proposed
solution addresses these concerns and ensures its security as well
as efficiency.

\bibliographystyle{unsrt}

\end{document}